\documentclass[conference]{IEEEtran}
\IEEEoverridecommandlockouts
\usepackage{cite}
\usepackage{amsmath,amssymb,amsfonts}
\usepackage{algorithmic}
\usepackage{graphicx}
\usepackage{textcomp}
\usepackage{xcolor}
\usepackage{todonotes}
\usepackage{csquotes}
\usepackage{hyperref}
\def\BibTeX{{\rm B\kern-.05em{\sc i\kern-.025em b}\kern-.08em
    T\kern-.1667em\lower.7ex\hbox{E}\kern-.125emX}}
\begin{document}

\title{College Spread of COVID-19 in Ohio\\

}

\makeatletter
\newcommand{\linebreakand}{%
  \end{@IEEEauthorhalign}
  \hfill\mbox{}\par
  \mbox{}\hfill\begin{@IEEEauthorhalign}
}
\makeatother

\author{\IEEEauthorblockN{Akinkunle Akinola and Kimberlyn Brooks}
\IEEEauthorblockA{\textit{Data Science Program} \\
\textit{Bowling Green State University}\\
Bowling Green, OH, USA \\
\{akinola,kbrooks\}@bgsu.edu}
\and
\IEEEauthorblockN{Akoh Atadoga, Vibhuti Chandna, and Robert C. Green II}
\IEEEauthorblockA{\textit{Department of Computer Science} \\
\textit{Bowling Green State University}\\
Bowling Green, OH, USA \\
\{atadoga,vchandn,greenr\}@bgsu.edu}
\and


}

\maketitle

\begin{abstract}
Cumulative COVID-19 case counts by county in Ohio were gathered and combined with population data from the Census Bureau and student enrollment by county from the Integrated Postsecondary Education Data System (IPEDS) for colleges or universities that compete in NCAA sports. Monthly median percent increases, monthly average percent increases, and average cases per 100k of COVID cases between counties with colleges and counties without colleges were calculated. The Wilcoxon test was used to determine if the samples were similar, and the analysis found the differences were statistically significant. Metro and non-metro groupings were added to the average cases per 100k to further subdivide the data set. Analysis found no statistically significant differences.
\end{abstract}

\begin{IEEEkeywords}
COVID-19, college, university, county, Ohio, test
\end{IEEEkeywords}

\section{Introduction}
In the spring of 2020, a global pandemic began to spread across the United States. Many colleges and universities quickly moved their courses online and sent students home to finish out the semester, including those in the state of Ohio. As it became more evident that COVID-19 was not going to be easily subdued, institutions had to come up with alternate plans for how to safely deliver educational experiences to students while protecting the health and well-being of students, faculty, and staff. Armed with a variety of instructional modes and guidelines from the Centers for Disease Control and Prevention (CDC)\cite{b17}, colleges and universities began making plans for students to return for a very different type of experience in the fall of 2020\cite{b3}\cite{b9}. As the pandemic continued through 2021, colleges and universities searched for ways to safely bring students back to campus for an in-person experience. This study analyzes community spread rates to understand the effects of these mitigation plans by comparing the median percent change, the average percent change, and the average cases per 100,000 people of COVID-19 in Ohio counties with and without NCAA colleges.

In order to perform the analysis, several sources of data needed to be considered. First and foremost COVID-19 case count data both at the daily case count level and the cumulative case count level by county was required. Daily case count data was obtained from the Ohio Department of Health\cite{b24} and was used to calculate the cumulative case counts. College data was gathered from the Integrated Postsecondary Education Data System (IPEDS)\cite{b15} and was used to determine whether the county contained a NCAA competing college. The National Center for Health Statistics\cite{b27} data was used to determine whether a county is classified as a metro or non-metro region by the Centers for Disease Control. US Census Bureau\cite{b0} data was also added to scale the data for the average cases per 100,000 data set.

In the rest of this article, a detailed distinction is made between the distribution of COVID-19 in college and non-college counties in Ohio. Section II explains the datasets used and describes the age distribution of COVID-19 in Ohio. Describing data analysis and hypothesis testing, Section III walks through the methodology. The results of the data visualization are addressed in Section IV and statistical inference from testing of hypotheses is given. The research conducted and its drawbacks are outlined in Section V and the future directions are provided.

\section{Background}

Throughout the pandemic, various studies have been conducted related to college campuses and the spread of COVID-19, with mixed results. Particularly concerning for colleges and universities was the positivity rate among persons 18-24. The positivity rate for this group was 14\%, which was the highest of any age bracket in September of 2020 \cite{b35}.  College campuses are at risk to develop an high incidence of COVID-19 and become superspreaders for neighboring communities \cite{b45}. Meanwhile, more than 80\% of our university county sample did not experience a significant case increase in Fall 2020 and there were no significant relationships between opening approaches and community transmission in both mask-required and non-mask-required states \cite{b34}. Referring to the same semester, fall of 2020, 

\begin{displayquote}
    County-level COVID-19 incidence decreased in much of the United States in late summer 2020. Comparing the 21 days before and after instruction start dates, university counties with in-person instruction experienced a 56\% increase in incidence and 30\% increase in hotspot occurrence as well as increases in COVID-19-related testing and test percentage positivity. Results from the unmatched analysis were consistent with those from the matched analysis. If percentage positivity had been stable or declining across the observation period, then efforts on the part of many colleges and universities to conduct or require testing before students’ return to campus and their ongoing surveillance efforts might explain an increase in case counts, as a result of increased case discovery. However, the concurrent increases in percentage positivity and in incidence in these counties suggest that higher levels of transmission, in addition to increased case discovery, occurred in these communities\cite{b34}.
\end{displayquote}
    
Yet, the \textit{New York Times} reported that, ``When many campuses reopened in the fall, outbreaks raced through dorms, infected hundreds of thousands of students and employees, and spread to the wider community. After students returned for the spring term this year, increased testing, social distancing rules and an improving national outlook helped curb the virus on many campuses''\cite{b1}.
The Center for Disease Control and Prevention (CDC) established and continually updated guidance for Institutions of Higher Education (IHE) to prevent transmission of COVID-19, although the guidance was considered supplemental and not intended to replace any federal, state, tribal, local, or territorial health and safety laws, rules, and regulations with which IHEs must comply \cite{b2}. The guidance included masking policies, isolation and quarantine recommendations, and testing/screening strategies. 
Even though there was guidance provided by the CDC and funding provided by the Coronavirus Aid, Relief, and Economic Security Act (CARES), college and university plans for instruction varied widely for the fall of 2020 and beyond. Six factors were statistically significant in determining whether instruction would be in-person, hybrid, or online. The six factors were: the financial strength of the institution (using average faculty salary as a surrogate), the selectivity (admission rate), whether the institution was public or private, the political affiliation of the state in which the institution was located, whether the institution was religiously affiliated, and the percentage of humanities degrees offered\cite{b43}. Similarly, the proportion of COVID-19 cases, deaths, and mask mandates had no association with colleges’ decisions to open their campuses. Rather, reopening decisions were driven largely by state and county politics as well as by budgetary concerns\cite{b42}. Ohio institutions vary widely with respect to the variables considered, so it is no surprise that the opening plans of the colleges and universities in Ohio varied. Instructional modes were also modified in response to positivity rates on campuses. Since the mode of instruction was a fluid variable, it was not considered as part of the following analysis.

\section{COVID-19 in Ohio}
An exploratory data analysis on the data sets obtained from the Ohio Department of Health\cite{b24} was covered. The Python pandas library was used to process and extract information from the available data set and appropriate graphs were created for better visualizations using matplotlib, and Seaborn libraries of Python.
\subsection{COVID-19 Spread in Ohio over time}

In Fig. 1, the x-axis represents the months from 2020-01 to 2022-02, and the y-axis represents the number of cases by onset date as of January 2020. On March 9, Governor Mike DeWine reported Ohio's first 3 cases in Cuyahoga County, and by December 1, the number of total cases had increased to 430,093 in Ohio\cite{b18}.

\begin{figure}[htbp]
\centerline{\includegraphics[width=1.0\columnwidth]{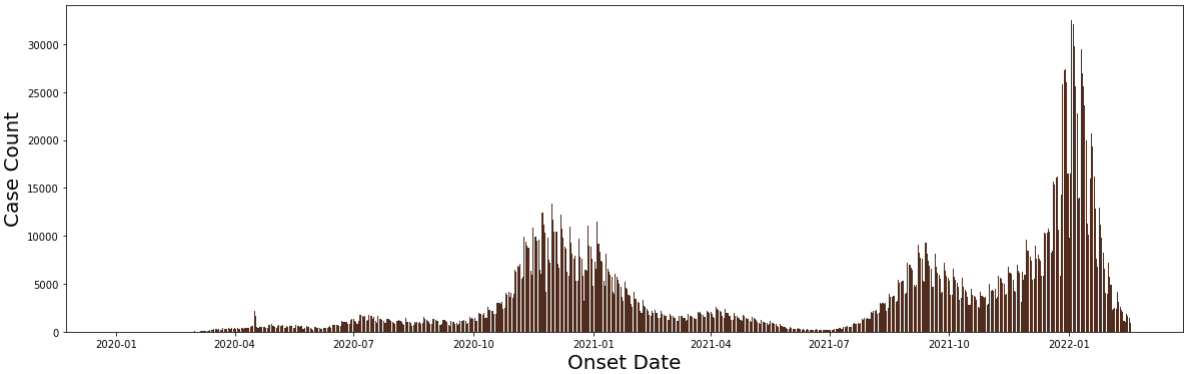}}
\caption{COVID-19 cases by onset date}
\label{fig}
\end{figure}

\subsection{Age and Gender distribution of COVID-19 cases in Ohio}
In the case of the COVID-19 pandemic, the importance of presenting epidemiological data by age and sex groups has been emphasized. Without this knowledge, the public is unable to make fully informed decisions regarding their own risk of disease, and reactions to public policy cannot be clearly tailored. Fig. 2 analyzes the spread of COVID-19 in Ohio to understand which age group is most affected. It appears that age group 20-29 is affected the most i.e. 19.7\% as compared to elderly who are more prone to disease, only account for 12.5\%. The 12.3\% are in between 0-19 years, 29\% in 30-49 years, and 27\% in the 50-69 age group. Fig. 3 shows the gender distribution of COVID-19 cases across different age groups. Women account for more COVID-19 positive cases as compared to men, independent of age, in the state of Ohio.

\begin{figure}[htbp]
\centerline{\includegraphics[width=1.0\columnwidth]{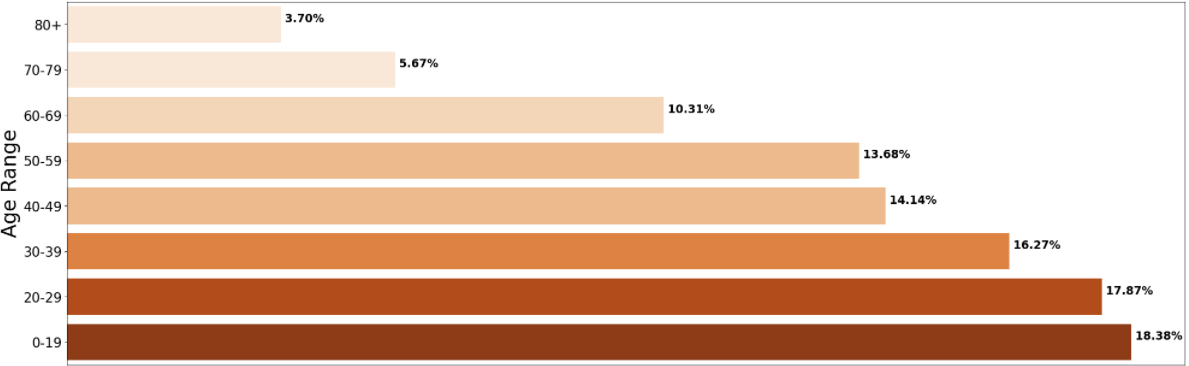}}
\caption{COVID-19 cases in Ohio distribution by age group}
\label{fig}
\end{figure}

\begin{figure}[htbp]
\centerline{\includegraphics[width=1.0\columnwidth]{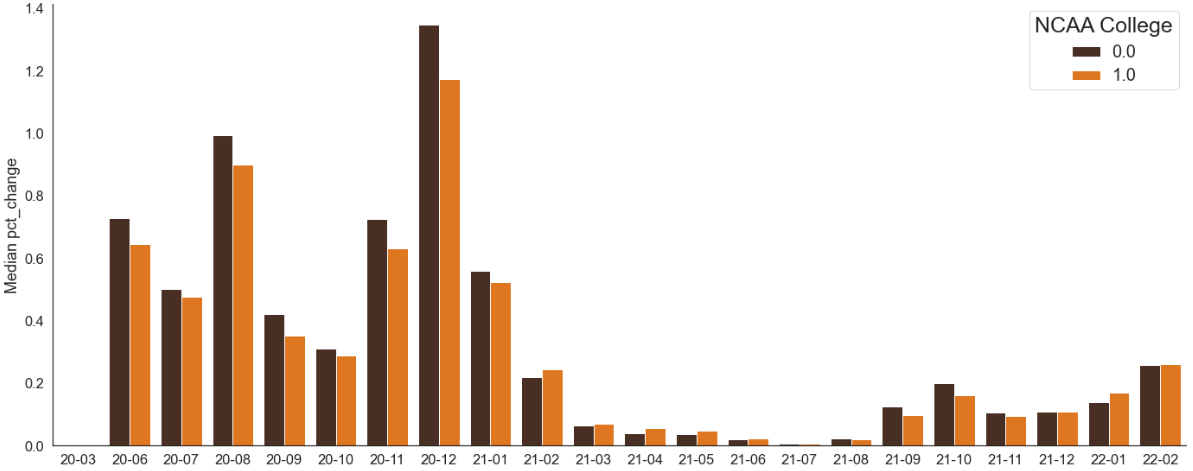}}
\caption{Monthly median percent change in COVID-19 cases in Ohio}
\label{fig}
\end{figure}

From the preliminary data analysis and visualization, it can be seen that thousands of new coronavirus cases continue to emerge in the age group 20-29, most of which represents the college student population. Governor Mike DeWine also mentioned that the sudden increase in the number of cases is connected to students returning to college campuses for the fall semester\cite{b21}.

\subsection{Other Pertinent Data}
Noting that the sudden increase in number of COVID-19 cases has been attributed to the opening of colleges and students returning to campuses, an impetus was provided to compare the COVID-19 spread in Ohio counties with and without NCAA affiliated colleges or universities. For this, the data of colleges including the county Federal Information Processing Standards (FIPS) code was obtained from the Integrated Postsecondary Education Data System (IPEDS)\cite{b15}. Data from the New York Times COVID-19 data repository\cite{b12} contained COVID-19 case counts by date, county, and the FIPS code. The 2020 county population was retrieved from the Census Bureau\cite{b0}. Lastly, data was obtained from the National Center for Health Statistics (NCHS)\cite{b27}  to obtain the Urban Rural Classification Scheme for counties. 

\section{Methodology}

\subsection{Data set cleaning}\label{AA}

There are many existing datasets tracking COVID-19 cases worldwide. Since this study is specific to Ohio, the Ohio Department of Health (ODH) COVID-19\cite{b24} was selected as the basis of the data used. The data on the ODH website is also updated regularly, which provided the ability to update the analysis over time. The dataset contains county, sex, age range, onset date, admission date, date of death, case count, hospitalized count, and death due to illness count by county of residence. The subject of interest in this study is case counts by age range and the monthly increase in case count by age range. A dataframe was constructed transforming the daily case counts into cumulative case counts by date and county. Since the fields sex, admission data, date of death, hospitalized count and death due to illness count were not needed, they were removed.

In order to make comparisons between counties that contained NCAA colleges and those that did not, data from the Integrated Postsecondary Education Data System (IPEDS)\cite{b15} database for 2020 was merged by county into the cumulative case counts by data and county (working) dataframe. The last classification that was added to the working dataframe is whether the county is classified as a metro or non-metro county. The National Century for Health Statistics (NCHS)\cite{b27} from the Center for Disease Control (CDC) dataset for the year 2013 was merged to the working dataframe which distinguished the classification of each Ohio county. The last data source needed to complete the study was the United States Census Bureau\cite{b15}. The 2020 population by county was merged into the working dataframe.

Before conducting the analysis, the data was trimmed to only include the first day of each month and the percent monthly change was calculated. Since the initial months with 0 cases showed a \texttt{NaN} in the percent change column, the NaN values were changed to 0. An additional column was added to the working dataframe to calculate the cases per 100k. The average and median percent change were calculated according to the classification of whether the county contained an NCAA college or not. The average and median percent change were also calculated by using four classifications, metro county with and without an NCAA college and non-metro county with and without an NCAA college. Outliers were found for the months of April and May of 2020 using the interquartile ranges and were omitted from the dataframe before the statistical analysis. No outliers were detected in the average cases per 100k field, so all data was considered for the analysis.

\subsection{Visualization}
After cleaning and processing the data set, multiple visualizations in forms of graphs were developed. Variables in the data set like the number of cases recorded per day, the amount of patients who were hospitalized, and the death counts due to COVID-19 for different age groups and gender were analyzed. After which the number of COVID-19 cases for every county was scaled to cases per thousand.

\subsection{Graphs}
Matplotlib and Seaborn were used to make multiple bar plots graphs to visualize the results. The following bar plots were constructed:

\begin{enumerate}
\item Fig. 1 showing the number of cases in Ohio by onset date
\item Fig. 2 showing the case count for different age groups 
\item Fig. 3 showing the median rate of change of COVID-19 cases by month
\item Fig. 4 showing the box plot of cases per 100k versus college and metro groupings
\item Fig. 5 showing the average case counts per 100k by metro/non-metro, with NCAA college/without NCAA college
\end{enumerate}

\subsection{Normality Test}
Understanding the importance of normal distribution is crucial as it is a fundamental assumption of many statistical approaches. When a statistical analysis is being conducted, it is vital to verify the most used distribution in statistical theory and applications. So, validating the assumption of normality is essential for the analysis\cite{b6}.
Amongst various normality tests, Shapiro Wilk, Jaque and Kolmogorov-Smirnov tests were used to check if the average cases per 100 thousand, average percentage change, and median percent change are normally distributed to be able to choose between parametric or non-parametric approach. Shapiro Wilk is among the most widely used of regression tests, with its value in the range of 0 and 1, which determines the normality/non-normality status of the data. The SW values contribute immensely to the P-value when the test is carried out\cite{b53}. On the other hand, Jarque and Bera's (JB ) test has optimum asymptotic power properties and good finite sample performance\cite{b54}. The JB tests for normality were obtained using the Lagrange multiplier procedure. Its statistics follow a chi-square distribution with 2 degrees of freedom. The hypothesis is as follows: The null hypothesis indicates that the skewness is 0 and kurtosis is 3. The null hypothesis of normality fails to reject if there are small values of skewness and kurtosis values close to or equal to 3\cite{b53}. In addition to the SW and JB tests, Kolmogorov-Smirnov, a highly regarded empirical distribution function (EDF) test used in testing the normality of data, is aimed to establish a comparison between EDF and the cumulative distribution function (CDF) of normal distribution\cite{b53}.
 It is recommended that the Shapiro Wilk test is the best to use whenever there is a sign that the distribution is asymmetric. JB's simplicity makes it useful for validating data's normality state\cite{b54}.

\begin{table*} [hbt!]
\centering
\addtolength{\tabcolsep}{0pt}
    \begin{tabular}{|p{3cm}|p{3cm}|p{2cm}|p{2cm}|p{2cm}|p{2cm}|p{2cm}|}
     
        \hline
        \textbf{Data} &  \textbf{Test} & \textbf{Significance}  & \textbf{P-value} & \textbf{Test Statistic} & \textbf{Conclusion}\\
        \hline
        Avg cases per 100k & Shapiro & 0.05 & 0.00 & 0.90 & Reject Null\\
        \hline
        Avg cases per 100k & Jarque & 0.05 & 0.19 & 3.31 & Reject Null\\
        \hline
        Avg cases per 100k & Kolmogorov-smirnov & 0.05 & 3.18e-158 & 0.90 & Reject Null\\
        \hline
        
        Avg pct change & Shapiro & 0.05  & 0.00 & 0.81 & Reject Null\\
        \hline
        Avg pct change & Jarque & 0.05  & 0.00 & 11.71 & Reject Null\\
        \hline
        Avg pct change & Kolmogorov-smirnov & 0.05  & 1.07e-10 & 0.50 & Reject Null\\
        \hline
        Median pct change & Shapiro & 0.05  & 0.00 & 0.82 & Reject Null\\
        \hline
        Median pct change & Jarque & 0.05  & 0.00 & 18.11 & Reject Null\\
        \hline
        Median pct change & Kolmogorov-smirnov & 0.05  & 1.07e-10 & 0.50 & Reject Null\\
        \hline
    
    \end{tabular}
    \caption{Results of normality tests}
\end{table*}

It was seen that they are not normally distributed, which infers that non-parametric tests will be adopted for this analysis.

\subsection{Non-parametric test}

In cases where data are skewed and not distributed normally, the non-parametric experiment is mostly adopted\cite{b30}. We considered two non-parametric tests known as the Kruskal Wallis test and the Wilcoxon Signed-Rank test since the data in this case is skewed.

The Kruskal Wallis test was used to compare the continuous outcome in greater than two independent samples which were the cases per 100 thousands of metro and non-metro cities while Wilcoxon Signed-Rank was used to compare the paired samples of median percentage change of NCAA colleges and non- NCAA colleges as well as the comparison of their average percent changes.   

\begin{table*}[hbt!]
\centering
\addtolength{\tabcolsep}{0pt}
    \begin{tabular}{|c|c|c|c|c|}
     
        \hline
        \textbf{Data} &  \textbf{Significance}  & \textbf{P-value} & \textbf{Test Statistic} & \textbf{Conclusion}\\
        \hline
        Avg cases per 100k & 0.05 & 0.80 & 0.99 & Fail to Reject Null\\
        \hline
    
    \end{tabular}
    \caption{Results of Kruskal Wallis test
}
\end{table*}

In Wilcoxon Signed-Rank test, the null hypothesis H0 says that the average percent change in cases in counties with colleges is less than or equal the average percent change in non-college counties \cite{b33}. Since our prior theory was that the rate of change of cases in college counties was lower than it was in non-college counties, a 1-sided test was performed. With the P-value being 0.249, the Wilcoxon test gives no evidence against the null hypothesis of higher average percentage change of cases in college counties. However, the null hypothesis is rejected in the comparison between the median percent changes and the average cases per 100 thousand with respective p-values of 0.019 and 0.000. 

\begin{table*}[hbt!]
\centering
\addtolength{\tabcolsep}{0pt}
    \begin{tabular}{|c|c|c|c|c|}
     
        \hline
        \textbf{Data} &  \textbf{Significance}  & \textbf{P-value} & \textbf{Test Statistic} & \textbf{Conclusion}\\
        \hline
        Avg cases per 100k & 0.05 & 0.00 & 269.00 & Reject Null\\
        \hline
        Avg pct change & 0.05  & 0.25 & 135.00 & Fail to reject Null\\
        \hline
        Median pct change & 0.05  & 0.02 & 175.00 & Reject Null\\
        \hline
    
    \end{tabular}
    \caption{Results of Wilcoxon Signed-Rank tests
}
\end{table*}
\begin{figure}[htbp]
\centerline{\includegraphics[width=1.0\columnwidth]{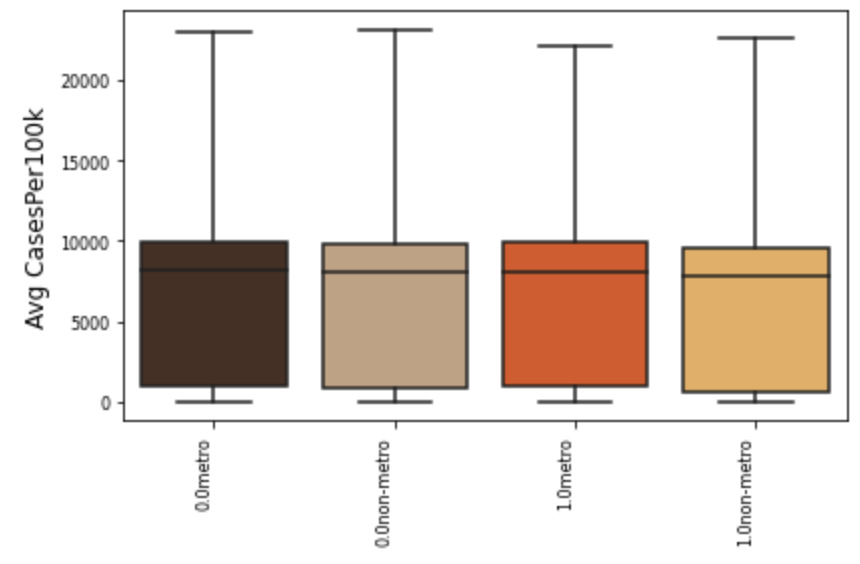}}
\caption{Box plot of Average cases per 100k vs College and metro grouping}
\label{fig}
\end{figure}

A further breakdown of the county data was done into metro and non-metro regions resulting in four groups- metro counties with and without NCAA colleges, and non-metro counties with and without NCAA colleges. This four group data was unbalanced because there was an uneven distribution of observations corresponding to each of the four groups- there were over double the number of observations for non-metro counties without NCAA colleges as compared to the metro counties with/without NCAA colleges, and there were almost four times the number of observations compared to the metro counties with NCAA colleges.  Hence, a Kruskal-Wallis analysis was performed on these four groupings.

The motivation of this test was to compare the four groups whether they are statistically different. This comparison yields a p-value of 0.80 which is obviously greater than the significant 0.05. The conclusion in this case is that the four groups are the same by failing to reject the null hypothesis. This implies that the mean of the cases per 100k are the same for the four groups and it is safe to say that whether the college is in the metro or non-metro, it does not impact COVID-19 cases.

\begin{figure}[htbp]
\centerline{\includegraphics[width=1.0\columnwidth]{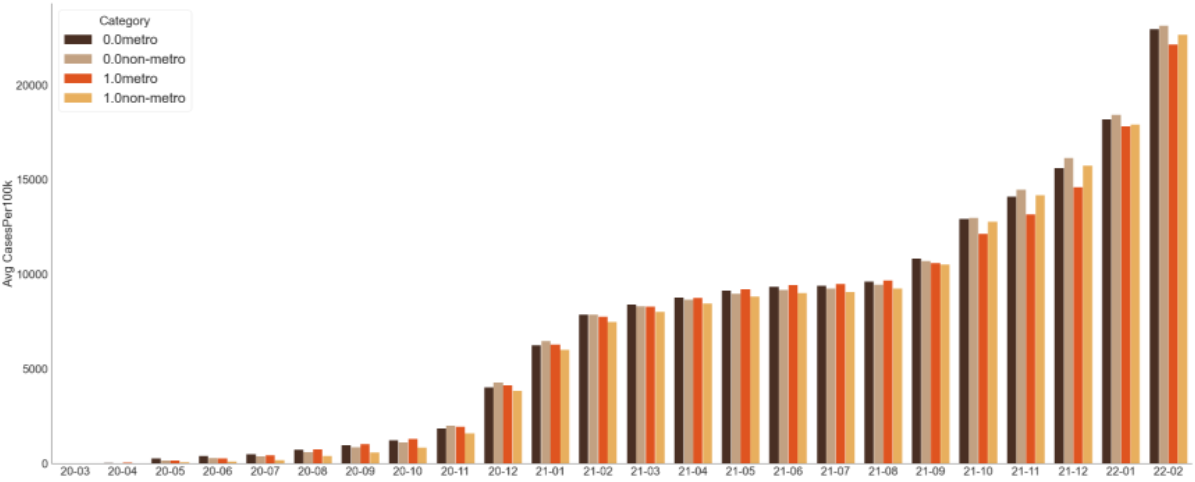}}
\caption{Average cases per 100k by college and metro grouping}
\label{fig}
\end{figure}

\section{Conclusion}
Since the COVID-19 pandemic is still currently ongoing and measures are still being taken to reduce the spread of the disease, the data set is incomplete and the analysis is far from final. The analysis this paper addresses is initial in its findings: In the state of Ohio, the median percent change  and the average cases per 100k in COVID-19 cases through February 1, 2022 is greater in counties with colleges than those without colleges. The average percent change in college counties is however less in counties with colleges. When further subdividing the cases per 100k by month and whether an NCAA college is present into metro and non-metro counties, a statistical comparison among the mean of the four groups resulted in the acceptance of the hypothesis that the means were the same.

A reflection on the mitigation strategies that colleges and universities used during the global pandemic is crucial for future planning. Infectious diseases is ranked sixth in the World Economic Forum Global Risks Perception Survey 2021-2022 \cite{b46}. It is also noted that 26.4\% of the countries surveyed indicate that infectious disease will become a critical threat to the world in 0-2 years \cite{b46}.

\begin{displayquote}
Since there are four endemic coronaviruses that circulate globally in humans, coronaviruses must have emerged and spread pandemically in the era prior to the recognition of viruses as human pathogens. The severe acute respiratory syndrome (SARS) coronavirus (SARS-CoV) emerged from an animal host, likely a civet cat, in 2002–2003, to cause a near-pandemic before disappearing in response to public health control measures. The related Middle East respiratory syndrome (MERS) coronavirus (MERS-CoV) emerged into humans from dromedary camels in 2012, but has since been transmitted inefficiently among humans \cite{b49}. COVID-19, recognized in late 2019, is but the latest example of an unexpected, novel, and devastating pandemic disease. One can conclude from this recent experience that we have entered a pandemic era \cite{b47} \cite{b50}.
\end{displayquote}
Certainly, planning for future pandemic events should be top of mind, particularly since "as climate change and urbanization progress, pandemics are likely to occur \cite{b52}.

\subsection{Limitations}\label{SCM}
There are many limitations to the analysis performed and many opportunities for further study:
\begin{itemize}
\item The pandemic is currently ongoing and the data is incomplete.
\item Reporting structure (which county the student case was reported) for student cases and number of student cases at each institution was not included.
\item An assumption was made that any increase is student cases, regardless of reporting structure, would increase the amount of community spread of COVID-19.
\item Only institutions that compete in NCAA sports were considered, but differences between them exist, i.e. small private college, large public university, etc.
\item Plans for students returning to campuses and modes of instruction varied widely\cite{b9} and were also not considered in this study.

\end{itemize}


\footnote{Source code available at: \href{https://gitlab.com/kbrooks47/covid-project/-/blob/main/DataPrepV17.ipynb}{Ohio COVID-19 Data Analysis}}

\vspace{12pt}

\end{document}